\documentclass[11pt]{article}
\usepackage{graphicx}
\usepackage{amsfonts}
\usepackage{amssymb}
\usepackage[centertags]{amsmath}
\usepackage{amsfonts,bm}
\usepackage{newlfont}
\usepackage{graphicx}
\usepackage{eurosym}
\usepackage{epstopdf}
\usepackage{caption}
\usepackage{subcaption}
\newtheorem{Theorem}{Theorem}

\newtheorem{lemma}[Theorem]{Lemma}
\newcommand{\E}{\mathbb{E}}
\newcommand{\var}{\mathrm{var}}

\newcommand{\R}{\mathbb{R}}
\newcommand{\C}{\mathbb{C}}
\newcommand{\e}{\mathrm{e}}

\newcommand{\eps}{\epsilon}

\newcommand{\cA}{{\cal A}}

\newcommand{\cC}{{\cal C}}

\newcommand{\cF}{{\cal F}}

\newcommand{\cL}{{\cal L}}

\newcommand{\cN}{{\cal N}}

\newcommand{\rmi}{\mathrm i}

\newcommand{\nt}{\noindent}

\newcommand{\lw}{\longrightarrow}

\begin{document}

\title{A moment matching method for option pricing under stochastic interest rates}

\author{F. Antonelli\footnote{DISIM, University of L'Aquila,
          \texttt{fabio.antonelli@univaq.it}}, A. Ramponi\footnote{Dept. Economics \& Finance, University of Roma - Tor Vergata, \texttt{alessandro.ramponi@uniroma2.it}}, S. Scarlatti\footnote{Dept. Enterprise Engineering, University of Roma - Tor Vergata, \texttt{sergio.scarlatti@uniroma2.it}}}

\maketitle

\begin{abstract}
In this paper we 
present a simple, but  new, approximation methodology for pricing a call option in a Black \& Scholes market characterized by stochastic interest rates. The method,  based on a straightforward Gaussian moment matching technique applied to a conditional Black \& Scholes formula, is quite general and it applies to various models, whether affine or not. 
To check its  accuracy  and computational time, we implement it for the CIR interest rate model correlated with the underlying, using the Monte Carlo simulations as a benchmark. The method's  performance turns out to be quite remarkable, even when  compared  with  analogous results obtained  by the affine approximation technique presented in \cite{GO} and by the expansion formula introduced in \cite{KK99}, as we show in the last section.

\medskip


\end{abstract}

\section{Introduction}

Since the appearance of the seminal Black \& Scholes/Merton option pricing fundamental formula, there has been an intensive effort to incorporate in the market model additional stochastic factors, such as the volatility and/or the interest rates, the latter already discussed by Merton himself in \cite{M73}.  Along the years,  a huge field of research developed,
leading to a very rich literature on stochastic volatility models, while  fewer papers aimed at the inclusion of a dynamic term structure into the valuation of derivatives,  e.g. \cite{Rab89}, \cite{AJ92}, \cite{STV93}, \cite{Rin95}, \cite{KK99},  \cite{DPS00}, \cite{RS09}.  

Nowadays, the improvement in the  performances of option pricing formulas  obtained by adding these risk factors is widely recognized in the empirical literature (see e.g \cite{BCC97}, \cite{BCC00}),  indeed in  (\cite{Kim02}) the author remarked that even including solely stochastic interest rates in the model does affect the pricing formula, especially for longer-dated options,  in a noticeable manner.

Of course this generalization implies a higher degree of mathematical complexity and the search for efficient pricing techniques, able to provide accurate answers in a short computational time (as opposed to Monte Carlo methods) has been relentless, even more so in  modern quantitative finance where a huge amount of data allows to consider strategies that call for on real-time model calibration.  Hence, computational efficiency has become one of the primary concerns of risk managers and this requirement essentially restricted the choice of models to the affine class (see \cite{DPS00}).

 Indeed, when the interest rates are modeled  in  a Gaussian processes framework, as in the  very popular Hull-White / Vasicek models, even analytical prices can be obtained. These models are appropriate for modeling periods that admit positive probability of negative rates, such as the current one, but this feature becomes a drawback in usual periods of positive rates. The most popular model used to avoid this drawback is  the Cox-Ingersoll-Ross (CIR)  one, which guarantees the rate's strict positivity under Feller's condition. Its popularity   comes from the fact that falls into the so called affine models, that can exploit a very efficient and fast Fourier transform technique  to price the bonds.
 
Unfortunately, the affinity of the model is lost when the interest rate is coupled, with correlation,  with a risky asset's dynamics, making the search for efficient approximations of risk-neutral pricing formulas very challenging. 

Here we present a simple, but  new, approximation methodology for pricing a European call option  in a market model given by a linear diffusion dynamics for the underlying (a Black \& Scholes (BS) framework) coupled with a  stochastic  short term risk-free rate. The problem is a classical one and the novelty lies on the fact that  we propose a quite straightforward moment matching (MM) technique, easy to implement and leading   to   very efficient approximations. 

In building our procedure a few  issues have to be addressed and we first provide, by appropriate conditioning,   a representation formula for the claim's price in terms of the BS formula, then we exploit a Gaussian approximation by properly matching the first two moments of the involved random variables, that allows to use the properties  of the Normal cumulative distribution function (c.d.f.)  (see Lemma (\ref{mean_N})).  
When applying the method to the affine models, we also employ  a change-of-numeraire technique (introducing the $T$-forward measure as in  \cite{BM13}) to  partially disentangle the contributions due to the underlying and to the interest rate, exploiting the explicit expressions of the bond's price in an affine framework. To keep computations as simple as possible, any time a quantity is computable, it is stored and treated as a constant in the sequel. This leads to an  efficient mixed use of the risk free probability and the $T$-forward measure to evaluate the separate quantities.

The paper is organized as follows. In Section 2 we derive a representation formula for the call option's  price in  Black \& Scholes market with stochastic risk-free short rates,  while in Section 3 the Moment Matching method is fully described. Finally, in Section 4   we restrict to the affine models and we apply our technique with a CIR interest rate. In the same Section, we briefly introduce other two techniques, the affine approximation, inspired by Grzelak and Oosterlee \cite{GO} and the expansion method proposed by Kim and Kunimoto \cite{KK99} alternative to  prices obtained by Monte Carlo simulations. Hence we run a numerical study comparing those methods with ours, using  Monte Carlo evaluation as a benchmark.

\section{The price of a European call in  the BS model with stochastic rates}

The underlying problem we are concerned with is the pricing of a European call option, whose payoff is given by the function
$f(x)=( \e^x- \e^\kappa)^+$ for some $\kappa\in \mathbb R$, when stochastic interest rates come into play.

Thus, given a  finite time interval $[0,T]$ and  a complete probability space $(\Omega, \cF, Q)$, endowed of a filtration $\{\cF_t\}_{\{t\in [0,T]\}}$ satisfying the ``usual hypotheses" (see  \cite{Pr}), the market model is defined by the log-price of a risky asset and a risk-free interest $(X_t,r_t)$, whose joint dynamic for any initial condition $(t,x,r) \in [0,T] \times \mathbb{R} \times \mathbb{R}$ and $\forall s \in [t,T]$  is given by
\begin{equation}\label{SDEsystem}
\!\!
\begin{cases}
X_s  =X_t +\! \int_t^s(r_v\!-\frac{\sigma^2}{2})dv+ \sigma \Big [\rho (B_s^1\!-B^1_t)+\sqrt{1\!-\!\rho^2}(B^2_s-B_t^2)\Big ], \,\,\, X_t=x\\
 r_s = r_t + \int_t^s\mu(v, r_v )dv+ \int_t^s\eta(v,r_v)dB^1_v,\quad r_t=r,
\end{cases}
\end{equation}
where $(B^1, B^2)$ is a two dimensional standard Brownian motion and $\rho\in (-1,1)$. 
Moreover, we assume that the deterministic functions $\mu(\cdot,\cdot)$ and $\eta(\cdot,\cdot)$ are in  a class that ensures the  existence and uniqueness of a strong solution of (\ref{SDEsystem}) (see e.g. \cite{KS}) and that $Q$ is some  risk neutral probability selected by the market. 

Under these assumptions, the pair $(X_t, r_t)$ is Markovian, whence 
the arbitrage-free option's  price is a deterministic function of the state variables,   given by
\begin{equation}\label{ufunct_rate}
u(t, x,r, T;\rho)=\E(\e^{-\int_t^Tr_s}(\e^{X_T(\rho)}-\e^{\kappa})^+ ds|X_t=x,r_t=r),
\end{equation}
provided that the coefficients $\mu $ and $\eta $ are chosen to guarantee the exponential integrability of $X_T$ and $\int_0^T |r_s| ds$. Here we wrote $X_T(\rho)$, to stress the   prices' dependence on the correlation parameter.

If $u(t, x,r,T;\rho)$ is regular enough in  $t,x,r$,  Feymann-Kac's formula  implies that it is a classical solution of the following two-dimensional parabolic problem
\begin{equation}\label{pde1}
\begin{cases}
 \frac{\partial u}{\partial t}+{\cL^{\rho}}u=0\\
u(T,x,r,T;\rho)=(\e^x-\e^{\kappa})^+,
\end{cases}
\end{equation}
where  $\cL^{\rho}= \cL^{\mathbf 0}+\cA$, with
\begin{eqnarray}
\label{op0}
\cL^{\mathbf 0}&:=&\biggl(\frac{\sigma^2}{2}\frac{\partial^2}{\partial x^2}+(r-\frac{\sigma^2}{2})\frac{\partial }{\partial x}-r\biggr)+\biggl(\frac{\eta^2(t,r)}{2}\frac{\partial^2}{\partial r^2}+\mu(t,r)\frac{\partial}{\partial r}\biggr)\\
\label{A_operator}
\cA &:=&
 \rho\sigma \eta(t,r)\frac{\partial^2}{\partial x\partial r}.
\end{eqnarray}
In what follows to keep notation easy,  we take $t=0$  and we  omit the dependence on $t$ in the pricing function. The general case may be readily obtained substituting  $T$ in the final formulas with the time to maturity $T-t$.

By conditioning internally with respect to $\cF^1_T=\sigma(\{B^1_s:0\leq s\leq T\})$, we have
\begin{equation}
u(x,r, T;\rho)=\E\left (\e^{-\int_0^T r_s ds} (\e^{X_T(\rho)}-\e^\kappa)^+\right) = \E\left (\e^{-\int_0^T r_s ds}\E\big ((\e^{X_T(\rho)}-\e^\kappa)^+|\cF^1_T\big )\right).
\end{equation}
But $X_T(\rho)\big |\cF^1_T\sim N(M_T, \Sigma_T)$, where 
$$
M_T=x + \int_0^T(r_s- \frac{\sigma^2}{2})ds + \sigma \rho B_T^1,\quad \text{and}\quad 
\Sigma^2_T = \sigma  (1-\rho^2)T.
$$
so we obtain
$$
\begin{aligned}
\E\left ((\e^{X_T(\rho)}-\e^\kappa)^+|\cF^1_T\right ) =& \e^{M_T+\frac{1}{2} \Sigma_T^2} \cN\left(\frac{M_T-\kappa+\Sigma_T^2}{\Sigma_T} \right) - \e^{\kappa} \cN\left(\frac{M_T-\kappa}{\Sigma} \right)\\
= &\e^{x + \int_0^T(r_s- \frac{\sigma^2}{2})ds + \sigma \rho B_T^1 +\frac{1}{2} \sigma^2 (1-\rho^2)T} \cN(d_1(\rho)) - \e^{\kappa} \cN(d_2(\rho)),
\end{aligned}
$$
where we  define
\begin{eqnarray}
d_1(\rho) & = & \frac{x-\kappa + \int_0^T r_s ds + \sigma \rho B^1_T + \frac{\sigma^2}{2} T - \sigma^2 \rho^2 T }{\sigma \sqrt{1-\rho^2} \sqrt{T}} \\
d_2(\rho) & = & \frac{x-\kappa + \int_0^T r_s ds + \sigma \rho B^1_T - \frac{\sigma^2}{2} T}{\sigma \sqrt{1-\rho^2} \sqrt{T}}
\end{eqnarray}
and $\cN$ denotes the cumulative distribution function of the standard Gaussian.
It is convenient to introduce the following notations
$$
\Lambda_T=\int_0^Tr_sds,\;\;\;\beta(T,\rho)=\frac{\rho}{(1-\rho^2)^{1/2}\sqrt{T}},\;\,\gamma(T,\rho)=\frac1{\sigma(1-\rho^2)^{1/2}\sqrt{T}}
$$
$$
\alpha_1(x,T,\rho)=\frac{x-\kappa +\frac{\sigma^2}{2}T-\sigma^2\rho^2T}{\sigma(1-\rho^2)^{1/2}\sqrt{T}},\;\;\;\alpha_2(x,T,\rho)=\frac{x-\kappa -\frac{\sigma^2}{2}T}{\sigma(1-\rho^2)^{1/2}\sqrt{T}},
$$
so that
\begin{equation}\label{unknown_RV}
d_i(x,T,\rho)=\alpha_i(x,T,\rho)+\beta(T,\rho)B_T^1+\gamma(T,\rho)\Lambda_T, \quad i=1,2.
\end{equation}
Setting $S_T=\e^{-\Lambda_T}$, we can finally write
\begin{equation} \label{uBS_rate}
u(x,r,T;\rho)= \e^x \e^{- \frac{\sigma^2 \rho^2}{2}  T}\E\left(\e^{\sigma \rho B_T^1 } \cN\big (d_1(\rho)\big )\right) - \e^{\kappa}\E\left(S_T \cN\big(d_2(\rho)\big)\right).
\end{equation}
In the forthcoming section we shall introduce the moment matching approximation procedure. 

\section{Option price approximation by moment matching}

The main idea of this section is to replace the r.v.'s $d_i(\rho)$, $i=1,2$, defined by  (\ref{unknown_RV}), with Gaussian r.v.'s $D_i(\rho)$ matching the first and second moments of $d_i(\rho)$.

\begin{figure}
\vspace{-5cm}
\hspace{-2.5cm}
\begin{subfigure}[b]{.6 \textwidth}
\includegraphics[width=12cm,height=17cm]{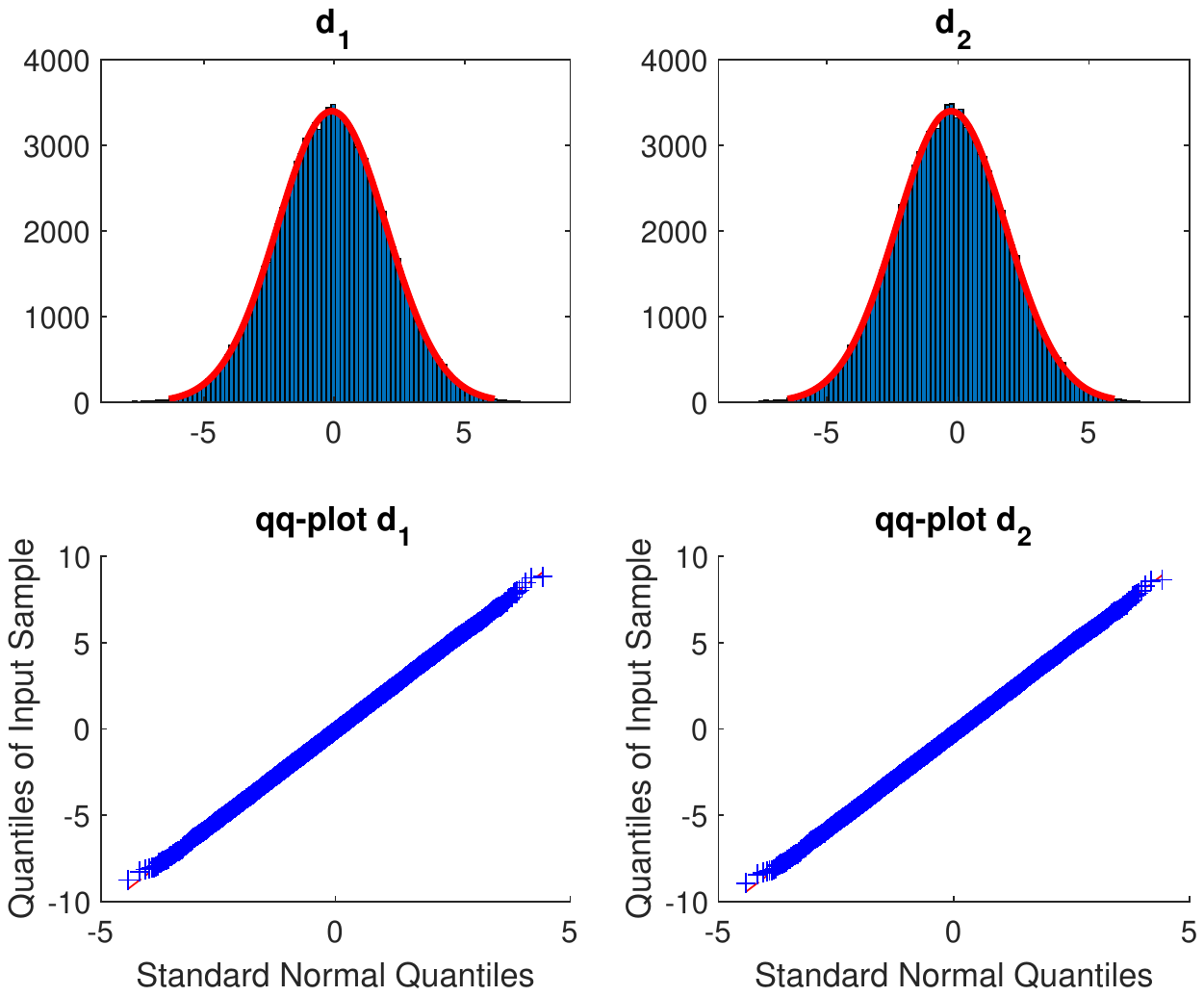}
\end{subfigure}
\hspace{-1cm}
\begin{subfigure}[b]{.6 \textwidth}
\includegraphics[width=12cm,height=17cm]{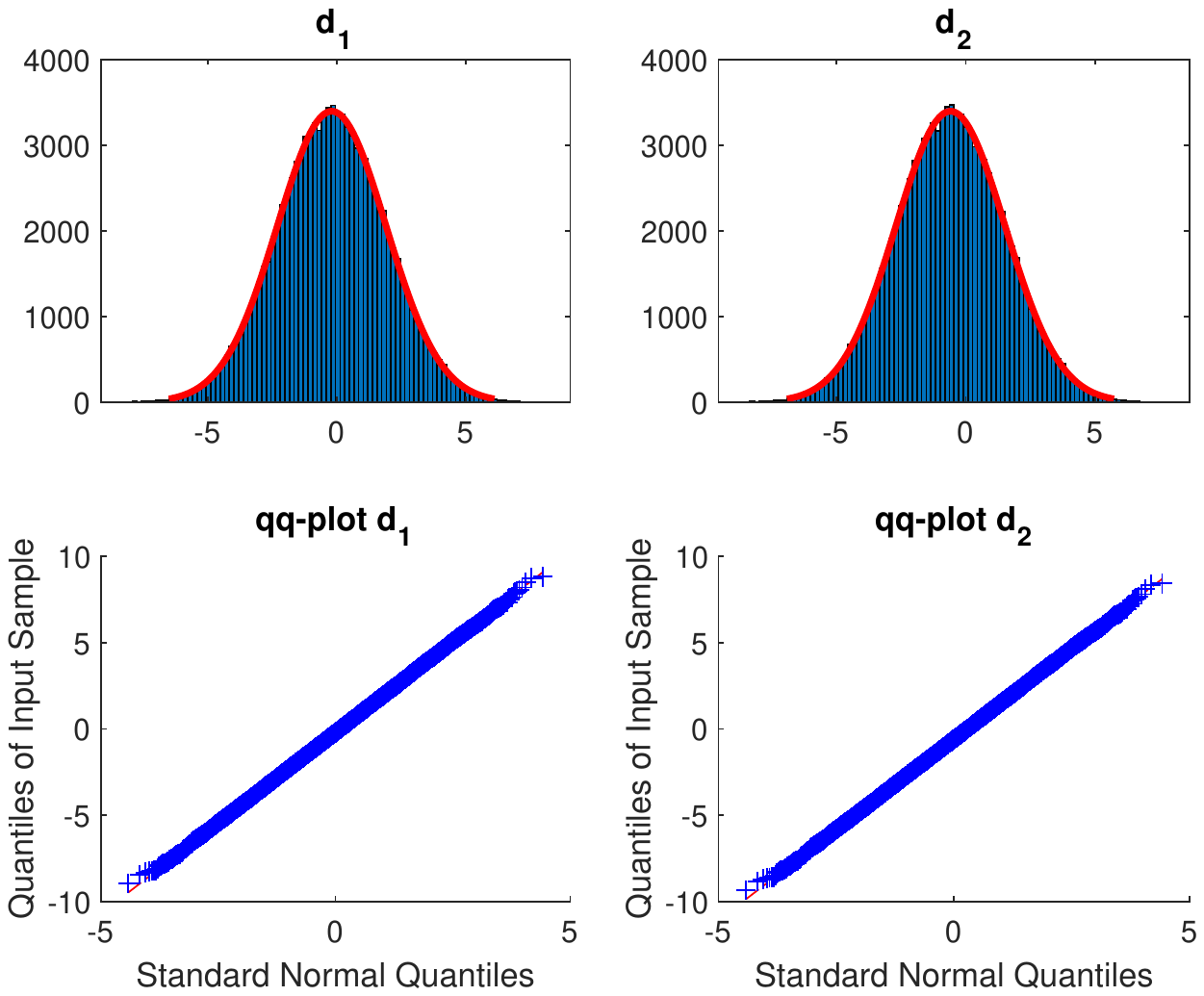}
\end{subfigure}
\vspace{-5.0cm}
\caption{\textit{The histograms of $d_1$ and $d_2$ for $\rho=0.3$,  $T=1$ (left) and $T=5$ (right), in comparison with the standard normal law (in red) and related qq-plot, CIR dynamic: $dr_t=\kappa(\theta-r_t)dt+\eta \sqrt{r_t} dB^1_t$.}}
\label{fig1}
\end{figure}

\begin{figure}
\vspace{-5cm}
\hspace{-2.5cm}
\begin{subfigure}[b]{.6 \textwidth}
\includegraphics[width=12cm,height=17cm]{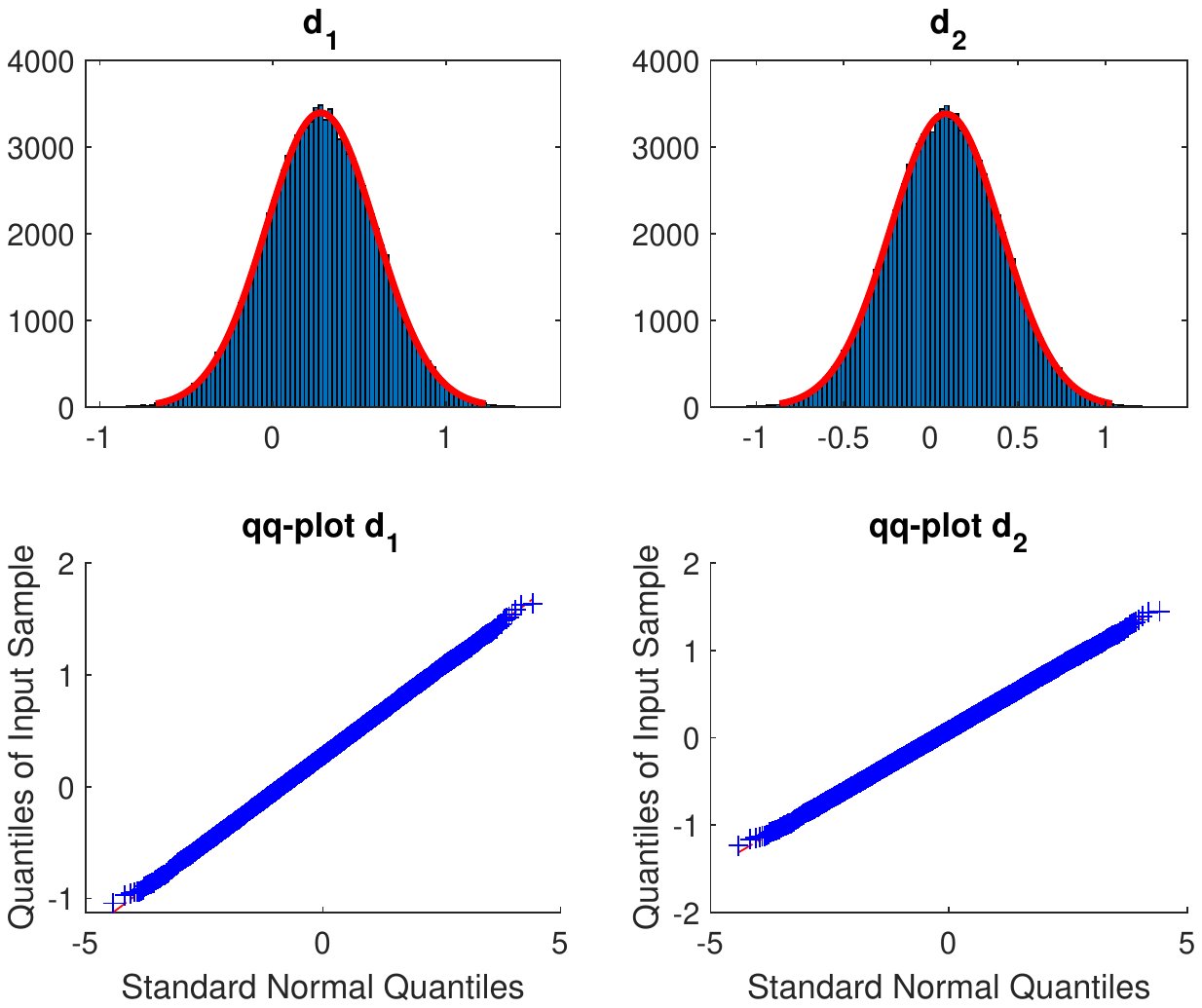}
\end{subfigure}
\hspace{-1cm}
\begin{subfigure}[b]{.6 \textwidth}
\includegraphics[width=12cm,height=17cm]{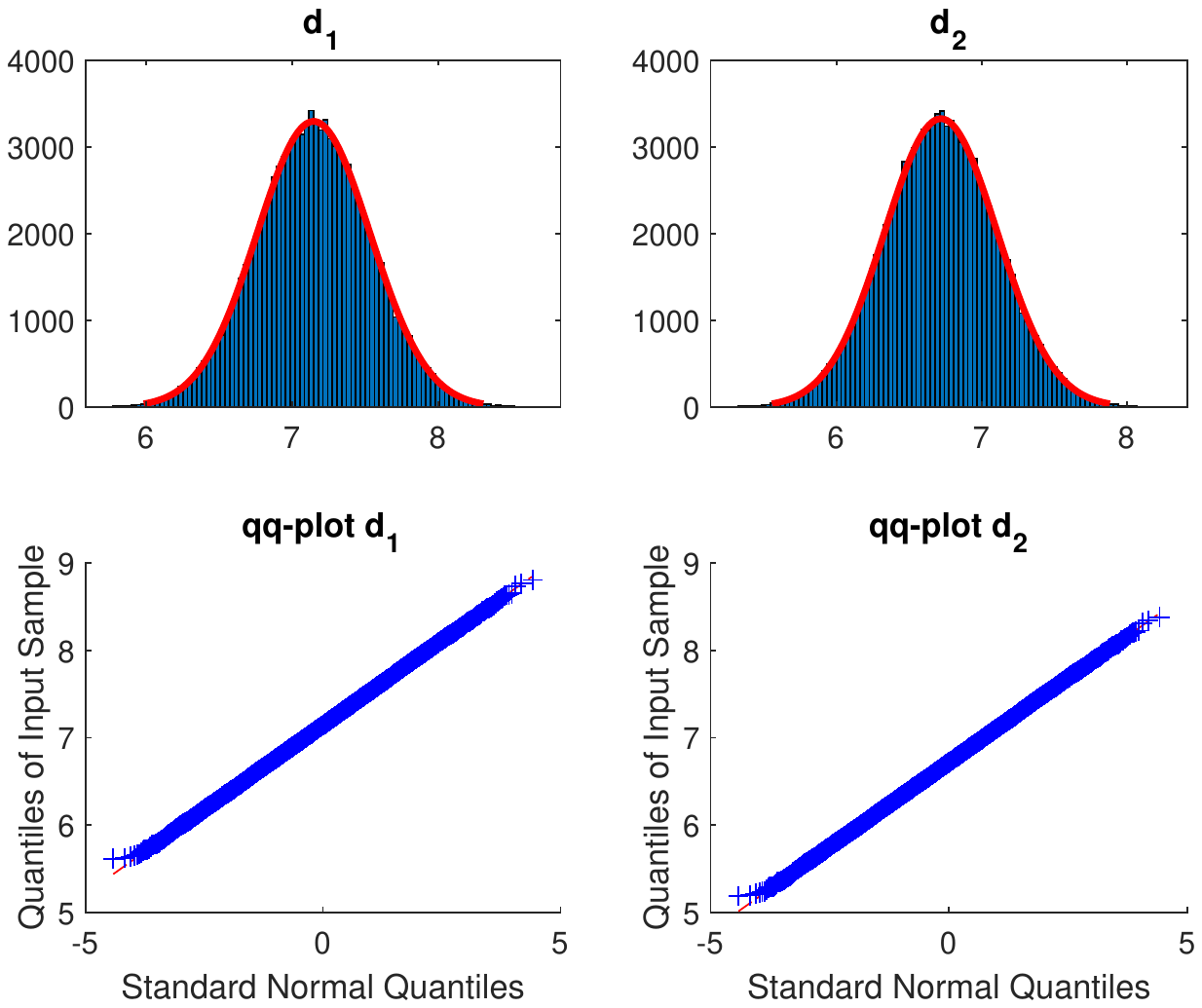}
\end{subfigure}
\vspace{-5.0cm}
\caption{\textit{The histograms of $d_1$ and $d_2$ for $\rho=0.3$,  $T=1$ (left) and $T=5$ (right), in comparison with the standard normal law (in red) and related qq-plot, Exponential Vasicek dynamic: $dr_t=r_t(\theta-a \ln(r_t))dt+\eta  r_t  dB^1_t$.}}
\label{fig2}
\end{figure}

\begin{figure}
\vspace{-5cm}
\hspace{-2.5cm}
\begin{subfigure}[b]{.6 \textwidth}
\includegraphics[width=12cm,height=17cm]{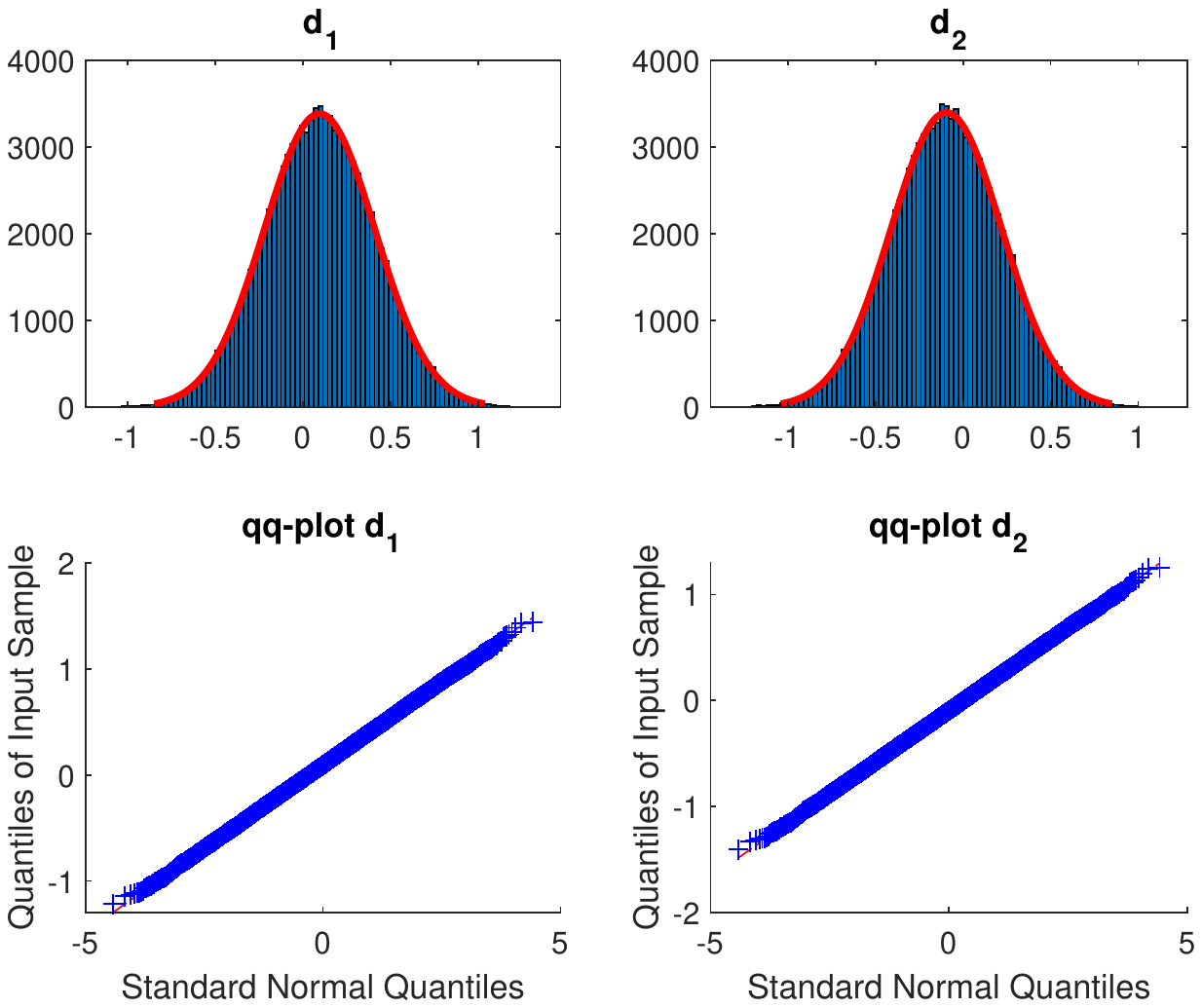}
\end{subfigure}
\hspace{-1cm}
\begin{subfigure}[b]{.6 \textwidth}
\includegraphics[width=12cm,height=17cm]{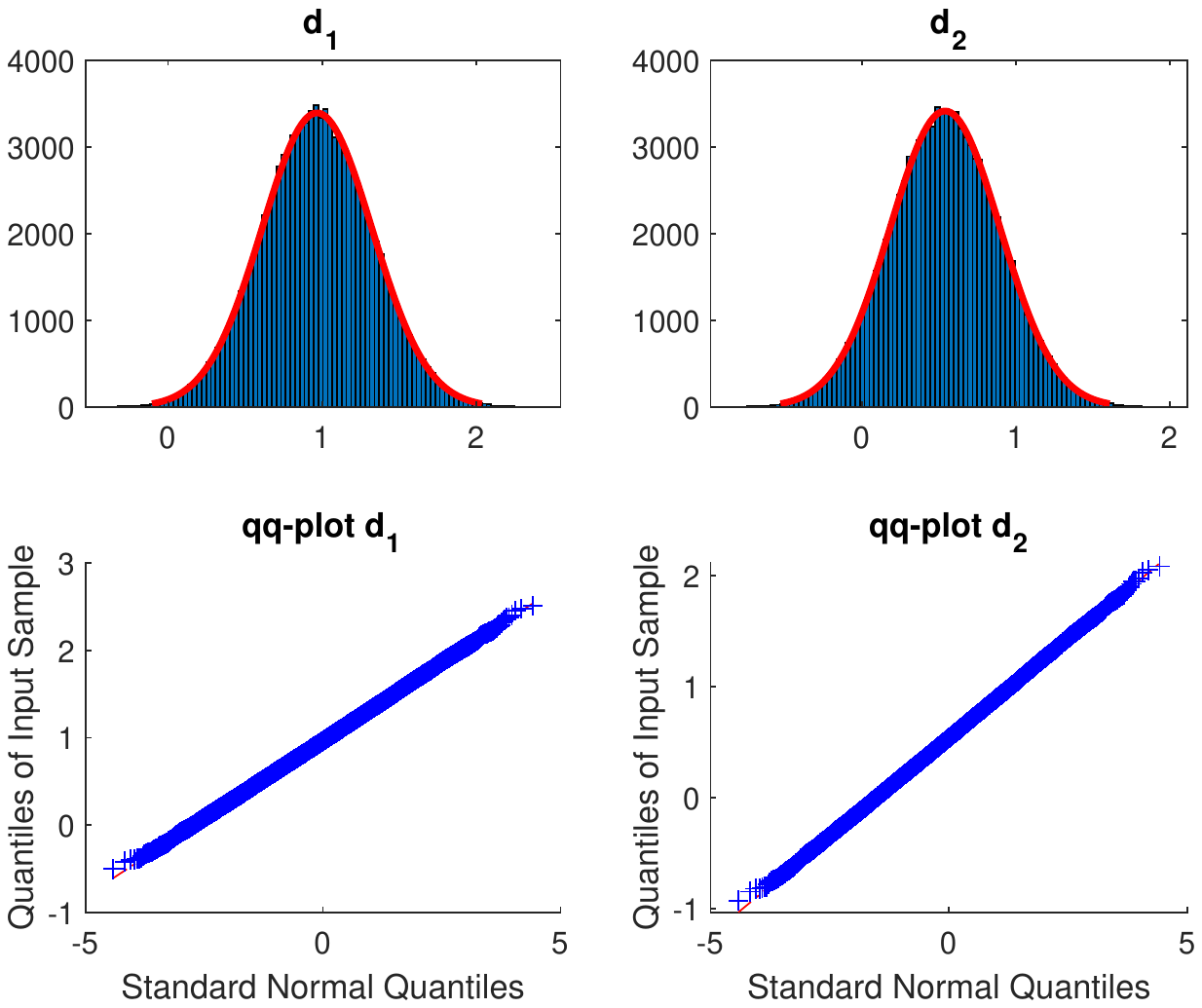}
\end{subfigure}
\vspace{-5.0cm}
\caption{\textit{The histograms of $d_1$ and $d_2$ for $\rho=0.3$,  $T=1$ (left) and $T=5$ (right), in comparison with the standard normal law (in red) and related qq-plot, Dothan dynamic: $dr_t=a r_t dt+\eta r_t  dB^1_t$.}}
\label{fig3}
\end{figure}

We  define
$$
D_i(\rho):= \alpha_i(x,T,\rho)+\hat \beta(T,\rho)B_T^1+\gamma(T,\rho)\E(\Lambda_T) ,\;\;\;i=1,2, 
$$
consequently
\begin{equation}
\label{mean}
\E\big (D_i(\rho)\big )= \alpha_i(x,T,\rho)+ \gamma(T,\rho)\E(\Lambda_T)= \E\big (d_i(\rho)\big )
\end{equation}
and  the new coefficient $\hat \beta >0$ is fixed such that
\begin{equation}
 \label{var_beta}
\begin{aligned}
\var\big (D_i(\rho)\big )=&
T\hat \beta^2(T,\rho)=\var(d_1(\rho))=\var(d_2(\rho))\\
=& \beta^2(T,\rho) T+ \gamma^2(T,\rho)\var(\Lambda_T) + 2 \beta(T,\rho) \gamma(T,\rho) \E(B_T^1 \Lambda_T)
\end{aligned}
\end{equation}
with
\begin{eqnarray} \label{var_lam}
\var(\Lambda_T) &= &\E\left( \Big(\int_0^T r_s ds\Big)^2\right)- \Big[\E\left (\int_0^T r_s ds\right)\Big ]^2, \\
\label{cross_mean}
 \E(B_T^1 \Lambda_T)&= &\E \left(B_T^1 \int_0^T r_s ds\right).
\end{eqnarray}
The moment matching method with Gaussian r.v's may be motivated by looking at the empirical distributional properties of the random variables $d_i$  in some well-known rate models: see as examples Figs (\ref{fig1}), (\ref{fig2}) and (\ref{fig3}).

We finally introduce a call price approximation
\begin{equation} \label{FG}
\begin{aligned}
u^{appr}(x,r,T;\rho):= &\e^x \e^{- \frac{1}{2} \sigma^2 \rho^2 T}\E\left(\e^{\sigma \rho B_T^1 } \cN\big(D_1(\rho)\big )\right) - \e^{\kappa}\E\left(S_T \cN\big (D_2(\rho)\big )\right)\\
 =: &\e^x \e^{- \frac{1}{2} \sigma^2 \rho^2 T}\ F(\rho)- \e^{\kappa} G(\rho).
 \end{aligned}
\end{equation}
The function $F$ can be evaluated in closed form by means of the following  
\begin{lemma}\label{mean_N}
Let $p\in \mathbb\R$ and $X\sim N(\mu,\nu^2)$, $(\mu,\nu)\in \mathbb R\times\mathbb R^+$,   then 
$$
\E(\e^{pX}\cN(X))=\e^{p\mu+\frac{(p\nu)^2}{2}}\cN\biggl(\frac{\mu+p\nu^2}{\sqrt{1+\nu^2}}\biggr).
$$ 
\end{lemma}

\nt
\begin{Proof} See \cite{ZA81} for $p=0$, the general case follows by a ``completing the squares" argument.\hfill $\square$
\end{Proof}

\medskip

Since
$$
B_T^1 = \big [D_1(\rho) - \alpha_1(x,T,\rho) -\gamma(T,\rho)\E(\Lambda_T)\big ] \hat \beta(T,\rho)^{-1},
$$
we may rewrite $F$ as
$$
\begin{aligned}
F(\rho)  = & \E\left(\e^{\sigma \rho (D_1(\rho) - \alpha_1(x,T,\rho) - \gamma(T,\rho) \E(\Lambda_T))\hat \beta(T,\rho)^{-1} } \cN(D_1(\rho))\right)\\
  = & \e^{- \sigma \rho [\alpha_1(x,T,\rho) +\gamma(T,\rho)\E(\Lambda_T)] \hat \beta(T,\rho)^{-1} } \E\left(\e^{\sigma \rho D_1(\rho) \hat \beta(T,\rho)^{-1}}\cN(D_1(\rho) )\right) \\
  = &  \e^{- \sigma \rho [\alpha_1(x,T,\rho) +\gamma(T,\rho)\E(\Lambda_T)] \hat \beta(T,\rho)^{-1} } \e^{\sigma \rho \E(D_1(\rho))\hat \beta(T,\rho)^{-1} +\frac{\hat \beta(T,\rho)^{-2} \sigma^2 \rho^2\var(D_1(\rho))}{2} } \\
    & \times  \cN\Big(\frac{\E(D_1(\rho))+\sigma \rho \var(D_1(\rho))\hat \beta(T,\rho)^{-1}}{\sqrt{1+\var(D_1(\rho))}}\Big ).
\end{aligned}
$$
From \eqref{mean} and \eqref{var_beta},   we may conclude
\begin{equation} \label{F_rho}
F(\rho)  =   \e^{\frac{\sigma^2 \rho^2 T}{2} } \cN \left(\frac{\alpha_1(x,T,\rho) + \sigma \rho \hat \beta(T,\rho) T+ \gamma(T,\rho)  \E(\Lambda_T)}{\sqrt{1+\hat \beta^2(T,\rho) T}} \right). 
\end{equation}
If $\E(\Lambda_T)$  and  \eqref{var_lam}, \eqref{cross_mean} can be computed, then $F$ is totally explicit. From now on, we denote $\lambda(T):=\E(\Lambda_T)$ to point out this is a known constant.

On the contrary, the function $G$ cannot be evaluated  in such a straightforward manner, as it involves a detailed knowledge of the joint distribution of $\Lambda_T $ and $B^1_T$ and not only of their moments and covariance. So, to represent $G$, we suggest applying a 
 a change-of-numeraire technique that allows us to exploit the bond pricing theory.
 
 Let us define
\begin{equation} \label{bond}
P(s,T):=\E \left((\e^{-\int_s^Tr_v dv}|\cF_s \right),
\end{equation}
the Zero Coupon Bond price. Again,  since $r.$ is a  Markov process, $P(s,T)$ is a deterministic function of the state variable, say $g(s,r_s)$, which we assume to be $\cC^{1,2}([0,T]\times \mathbb R^+)$. For $0\le s\leq T$,  we define the $\cF_s$-martingale (we remark that is a true martingale thanks to the exponential integrability of $\Lambda_T$)
\begin{equation}
\label{Girmg}
L_s= \frac{\E(\e^{-\int_0^T r_v dv}|\cF_s)}{P(0,T)}= S_s\frac{P(s,T)}{P(0,T)} = S_s\frac{g(s,r_s)}{g(0,r)}, \quad r_0=r.
\end{equation}
By applying It\^o's formula, we have the dynamic of $L$
$$ 
\begin{aligned}
dL_s=&\frac{S_s}{g(0,r)}\Big [ \frac{\partial g}{\partial t}(s,r_s)+ \frac 12  \eta^2(s,r_s) \frac{\partial^2 g}{\partial r^2} (s,r_s)+ \mu (s,r_s) \frac{\partial g}{\partial r} (s,r_s)-r_s  g(s,r_s)\Big  ]ds\\
+ &\frac{S_s}{g(0,r)}\eta(s,r_s) \frac{\partial g}{\partial r} (s,r_s)dB^1_s=\frac{S_s}{g(0,r)}\eta(s,r_s) \frac{\partial g}{\partial r}(s,r_s) dB^1_s, \quad L_0=1
\end{aligned}
$$
and we may define the $T$-forward measure on every $A\in \cF$ by $Q^T(A):=\E(L_T1_A)$ (see  \cite{B} for the method and \cite{BR18} for a similar application).  Under $Q^T$, we get
\begin{equation}\label{newF}
G(\rho)=\E\left(S_T\cN(D_2(\rho))\right)=P(0,T)\E^{Q^T}\left(\cN(D_2(\rho))\right),
\end{equation}
and by Girsanov theorem, by setting
$$
\xi_s:=\int_0^s \frac {\eta(v,r_v) }{g(v,r_v)} \frac{\partial g }{\partial r}(v,r_v) dv,
$$
we have that the process 
$
\tilde B^1_s:= B^1_s - \xi_s
$
is a $Q^T-$Brownian motion.  When choosing an interest rate model that allows an explicit expression of the bond's price,
$\E^{Q^T}\left(\cN(D_2(\rho))\right)$ will be the last quantity to compute. Under  $Q^T$,   $D_2(\rho)$ has the expression
$$
D_2(\rho)=\alpha_2(x,T,\rho) + \xi_T \hat\beta(T,\rho)+\hat\beta(T,\rho) \tilde B_T^1 + \gamma(T,\rho) \lambda(T),
$$
whence its distribution is no longer known.

To compute the final expectation, we replace $D_2(\rho)$ by the r.v.
$$
\bar D_2(\rho):=\alpha_2(x,T,\rho) + \E(\xi_T) \hat\beta(T,\rho)+\hat\beta(T,\rho) \tilde B_T^1 + \gamma(T,\rho) \lambda(T),
$$ 
where we are taking  $\eps(T):= \E(\xi_T) $ under the original probability $Q$, so that $\bar D_2(\rho)$ is a Gaussian r.v. and we may apply Lemma 1 once again to obtain 
$$
\E^{Q^T}(\cN(\bar D_2(\rho))) = \cN \left( \frac{\E^{Q^T}(\bar D_2(\rho))}{\sqrt{1+\var^{Q^T}(\bar D_2(\rho))}} \right),
$$
with 
$$
\E^{Q^T}(\bar D_2(\rho))=\alpha_2(x,T,\rho) +\eps(T)\hat\beta(T,\rho) + \gamma(T,\rho) \lambda(T), \ \ \var^{Q^T}(\bar D_2(\rho))=\hat\beta^2(T,\rho) T.
$$
Hence we shall denote by
$$
\bar G(\rho):=P(0,T)\E^{Q^T}(\cN(\bar D_2(\rho))) 
$$
the approximation of $G(\rho)$ and we may define the final  approximation of the call option price $u(x,r,T;\rho)$ as
\begin{eqnarray}
\nonumber& &\bar u(x,r,T;\rho) :=  \e^{x- \frac{1}{2} \sigma^2 \rho^2 T}F(\rho) - \e^{\kappa}\bar G(\rho) \nonumber \\
 \label{approx}&= &\e^x \cN \left(\frac{\alpha_1(x,T,\rho)+ \sigma \rho \hat \beta(T,\rho) T+ \gamma(T,\rho) \lambda(T) }{\sqrt{1+\hat \beta^2(T,\rho) T}}\right)\\
\nonumber &- & \e^{\kappa} P(0,T) \cN \left( \frac{\alpha_2(x,T,\rho) + \eps(T) \hat\beta(T,\rho) +\gamma(T,\rho) \lambda(T)}{\sqrt{1+\hat \beta^2(T,\rho) T}} \right). 
\end{eqnarray}
As a conclusion, we summarize the key requirements to make the approximation (\ref{approx}) explicitly computable and hopefully efficient
\begin{enumerate}
\item the  distributions of $d_i(\rho), i=1,2$ should be close to a  Gaussian distribution;
\item the bond price $P(t,T)$ should be theoretically computable. Moreover one can exploit  the observed (today) bond price for  $P(0,T)$ in \eqref{Girmg} and for calibration purposes; 
\item the quantities $\E(\Lambda_T)$, $\var(\Lambda_T)$ and $\E(\Lambda_TB^1_T)$  and/or their approximations, should be easily computable;
\item the change of numeraire technique (Girsanov's theorem) should be applicable.

\end{enumerate}

The performance of this approximation needs to be compared with Monte-Carlo simulated prices and then with other methods present in the literature. This will be done in the next section.

\section{Numerics and comparison with other methodologies}

In this section we employ  an affine model for the interest rate. This choice provides an explicit expression for the the ZCB's price \eqref{bond}. So our market model is given by
\begin{equation}\label{SDEsystem_affine}
\begin{aligned}
X_s  =&X_t +\!\! \int_t^s(r_v- \frac{\sigma^2}{2})dv+ \sigma \Big [\rho (B_s^1\!-B^1_t)+\sqrt{1\!-\!\rho^2}(B^2_s-B_t^2)\Big ], \quad X_t =x\\
r_s = &r_t +\!\! \int_t^s[a(v) r_v + b(v)]dv+ \int_t^s[c(v) r_v + d(v)]^{1/2}dB^1_v,\quad r_t=r,
\end{aligned}
\end{equation}
where $a,b,c,d:[0,T]\lw \mathbb R$ are bounded functions.
In this framework, we have for $r_t=r$
$$
P(t,T)= g(t,r)=A(t,T)\e^{-r B(t,T)},
$$
for suitable deterministic functions $A(\cdot,T)$ and $B(\cdot,T)$. Two very classical models fall into this setting
$$
\begin{aligned}
\text {(Vasicek) }\quad &a(v) =  -\gamma,\quad b(v) = \gamma \theta, \quad c(v)=0, \quad d(v) = \eta^2\\
\text {(CIR) }\quad &a(v) =  -\gamma,\quad b(v) = \gamma \theta, \quad c(v)= \eta^2, \quad d(v) = 0 
\end{aligned}\quad \gamma, \theta>0,
$$
for which $A(t,T)$ and $B(t,T)$ are explicitly known (\cite{BM13}), the same being true also for the Hull-White / Vasicek and Hull-White / CIR models, considering time dependent coefficients.

The functions $A$ and $B$ are usually characterized by the solution of a Riccati system of ODE's. Unfortunately, when in presence of correlation, the same procedure cannot be applied to the pair $(X.,r.)$, since its diffusion matrix
\begin{equation}\label{no_affine}
\sigma(v,x,r)\sigma(v,x,r)^T=
\begin{pmatrix}
\sigma^2 & \rho\sigma [c(v) r + d(v)]^{1/2}\\
\rho\sigma [c(v) r + d(v)]^{1/2}& c(v) r + d(v)
\end{pmatrix}
\end{equation}
may haves entries which are non-linear in the state variables, so that the joint diffusion is no longer affine, as it happens for the CIR model.
Hence, in this context it makes sense to apply the approximation procedure presented in the previous section. As before we consider $t=0$.

In this case  (see e.g. \cite{BM13}), setting $ \delta=\sqrt{\gamma^2+2\eta^2}$, we have 
$$
A(0,T)=\e^{\frac {2\gamma\theta}{\eta^2}} 
\frac{2\delta \e^{\gamma+\delta T}}{\delta-\gamma+(\delta+\gamma)\e^{\delta T}},
\quad 
B(0,T)=\frac{2(\e^{\delta T}-1)}{\delta-\gamma+(\delta+\gamma)\e^{\delta T}},
$$   
and let us proceed to the computation of $\E(\Lambda_T)$, $\var(\Lambda_T)$ and $\E(\Lambda_TB^1_T)$.

\begin {enumerate}

\item \textbf{Computation of $\E(\Lambda_T) $}.
It is straightforward to see
$$
\E(\Lambda_T) = \int_0^T\E(r_s) ds=\int_0^T  \big [(r_0-\theta)\e^{-\gamma s}+ \theta\big ] ds= \theta T +(r_0-\theta)\frac {1- \e^{-\gamma T}}\gamma.
$$

\item\textbf{Computation of $\var(\Lambda_T) $.} Taking into account the first point, we only have to compute the second moment
$$
\begin{aligned}
\E\left(\Big (\!\! \int_0^T \!\!\!r_s ds\Big )^2\right)=& \E\left( \int_0^T\!\!\!\int_0^T \!\!\!r_s r_vds dv\right)= \int_0^T\!\!\!\int_0^T\E( r_s r_v) ds dv\\
=&\int_0^T\!\!\!\int_0^t\E( r_s r_v) ds dv+\int_0^T\!\!\!\int_t^T\E( r_s r_v) ds dv\\
=&\int_0^T\!\!\!\int_0^t\!\!\!\E( r_s r_v) ds dv+\!\!
\int_0^T\!\!\!\int_0^s\!\!\!\E( r_s r_v)  dv ds
= 2\!\!\int_0^T\!\!\!\int_0^s\!\!\!\E( r_s r_v) dv ds.
\end{aligned}
$$
By  the independence of the increments of the process $r$,  for $v<s$ we have
$$
\begin{aligned}
&\E(  r_sr_v) =\E\left(( r_s-  r_v) r_v+  r^2_v \right)=\E(  r_s-  r_v) \E( r_v) + \E( r_v^2)\\
=&
\E\Big (\theta (s-v) +(r_v-\theta)\frac {1- \e^{-\gamma (s-v)}}\gamma\Big )\E( r_v)+ \E( r_v^2)\\
=&\theta\Big [ (s-v)- \frac {1- \e^{-\gamma (s-v)}}\gamma\Big ]\E( r_v)+\frac {1- \e^{-\gamma (s-v)}}\gamma[\E( r_v)]^2+  \text{var} (r_v)+ [\E( r_v)]^2
\end{aligned}
$$
Since $\var(r_s) = r_0 \frac{\eta^2}{\gamma}(\e^{-\gamma s}-\e^{-2 \gamma s})+\frac{\theta \eta^2}{2 \gamma}(1-\e^{-\gamma s})^2$, all the integrals appearing in the second moment can be calculated analytically. 

\item \textbf {Computation of $\E(B_T^1 \Lambda_T)$}. By  It\^o's integration-by-parts formula, we get
$$
\E(B_T^1 \Lambda_T)= \int_0^T \E(B_s^1 r_s ) ds
$$
and again by integration by parts we have
$$
\begin{aligned}B^1_s r_s =&\int_0^s B^1_v dr_v+\int_0^s r_v dB^1_v  + \langle B^1, r\rangle_s\\
=& \int_0^s B^1_v \gamma (\theta-r_v) dt+ \eta \int_0^s B^1_t \sqrt{r_v} dB^1_v + \int_0^s r_v dB^1_v+ \eta \int_0^s \sqrt{r_v} dv,
\end{aligned}
$$
so that
$$
\E(B^1_s r_s ) = -\gamma\int_0^s \E(B^1_v r_v) dv  +  \eta \int_0^s \E(\sqrt{r_v}) dv.
$$
Solving this linear ODE for $h(s):=\E(B^1_s r_s)$, since $h(0)=0$, we obtain
$$
\begin{aligned}
\E(B^1_s r_s) =& \eta\int_0^s \e^{-\gamma(s- v)} \E(\sqrt{r_v}) dv,\\
\E(B_T^1 \Lambda_T)=& \eta \int_0^T \int_0^s \e^{-\gamma (s- v)} \E(\sqrt{r_v}) dv ds.
\end{aligned}
$$
Thus the final crucial point is computing $ \E(\sqrt{r_v})$, which is rather delicate (see \cite{Dufr}). No explicit expression can be provided and we employ the approximation proposed in \cite{GO}, that we are going to present in the next subsection
\begin{equation} \label{mean_sqrt}
\E(\sqrt{r_v}) \approx a + b \e^{-c v}
\end{equation}
where the parameters $a, b$ and $c$ are obtained by an ad hoc matching procedure, which proved to be numerically very efficient.
\end{enumerate}
Finally, given the above three points, $\hat \beta(T,\rho)$ is easily computed from (\ref{var_beta}).

As a last step, we have to approximate $\E(\xi_T)$, which is readily done, given  the last remark, since
$$
\begin{aligned}
\E(\xi_T) = &\E\Big (- \eta \int_0^T\!\!\!B(s,T)  \sqrt{r_s} ds\Big )= \E\Big (- \eta \int_0^TB(s,T)  \E\big (\sqrt{r_s}\big ) ds\Big )
\\
\approx& - \eta \int_0^TB(s,T) (a + b \e^{-c s})  ds.
\end{aligned}
$$

\subsection{The Grzelak-Oosterlee (GO) approximation}
Here and in the next subsection, for completeness, we briefly describe the two approximation techniques, we are going to compare with.

The GO approximation consists simply in modifying the $\cA$ operator given in (\ref{A_operator}) by replacing the state variable in the coefficient with a constant, namely we define
$$
\cA^{GO}u (s,x,r):=\rho\sigma\E( \eta(s, r_ts)\frac{\partial^2u}{\partial x\partial r}.
$$
In the case of the CIR model, $\eta$ is time-homogeneous and this operator becomes
$$
\cA^{GO}u (s,x,r):=\rho\sigma \eta \E(\sqrt{r_s}))\frac{\partial^2u}{\partial x\partial r}\approx \rho\sigma \eta (a + b \e^{-c s}) \frac{\partial^2u}{\partial x\partial r}
$$
Once this replacement has been made then the Fourier transform methods apply, hence it is possible to compute approximated prices of the call option. We shall denote this approximation by $u^{GO}(t,x,r,T;\rho)$. To evaluate the accuracy of this approximation a comparison with the prices of the (non-affine) true model, obtained by MC simulations, must be performed.

Once again we specialize the formulas  for $t=0$ for a direct comparison with our results, so $X_0=x$ and $r_0=r$.

The discounted transform, for $\zeta\in \C$,  (see \cite{DPS00}) for the affine approximation is  
$$
\phi(\zeta, x, r, T) := \E\left(\e^{-\int_0^T r_s ds} \e^{\zeta X_T}\right ) = \e^{A(\zeta,T)+B(\zeta,T)x + C(\zeta,T)r},
$$
where the functions $A,B,C$ satisfy a system of solvable ODE's,  that give 
$$
\begin{aligned}
B(\zeta,T) =& \zeta,\\
C(\zeta,T) =& \frac{1-\e^{-d T}}{\eta^2(1-g \e^{-dT})}, \ \ d=\sqrt{\gamma^2+2 \eta^2(1-\zeta)}, \ \ g=\frac{\gamma-d}{\gamma+d},\\
A(\zeta,T) = & -\frac{\sigma^2}{2}T \zeta (1+\zeta) + \frac{\gamma-d}\eta\int_0^T\Big [\frac{\gamma \theta}{\eta} +  \rho\sigma \zeta \overline{r^{sq}_s}\Big ]\frac{1-\e^{-d s}}{(1-g \e^{-d s})} ds,
\end{aligned}
$$
where $\overline r^{sq}_s= \E(\sqrt {r_s})$ is approximated as in (\ref{mean_sqrt}).

Finally, by L\'evy inversion formula as in \cite{DPS00} or Fourier inversion as in \cite{Lee04}, one gets an integral representation for the price function: in our implementation we use the Fourier inversion
\begin{equation}
u^{GO}(x,r,T;\rho) = \frac{\e^{\nu \gamma}}{\pi} \int_0^{+\infty} \mathcal{R}\left(\frac{\e^{-\rmi \zeta \gamma}}{\nu^2-\nu - \zeta^2+\rmi \zeta(1-2\nu)} \phi(\zeta, x, r,T)  \right) d\zeta,
\end{equation}  
where $\nu <0$ is a dumping factor and $\mathcal R(z)$ is the real part  for $z \in \C$.  

\subsection{The Kim-Kunimoto (KK) approximation}

Kim and Kunimoto, in \cite{KK99}, consider a Taylor expansion of the process  $r_s$ in powers of $\eta$ around $\eta=0$. Considering the first order polynomial and setting $\varphi(s)=r \exp(-\gamma s)+\theta(1- \exp(-\gamma s))$, they obtain
\begin{equation}
\label{KK1}
r_s=\varphi(s)+\eta \int_0^s \e^{-\gamma(s-v)}\sqrt{\varphi(v)}(\rho dB^1_v+\sqrt{1-\rho^2}dB^2_v)+o(\eta).
\end{equation}
Inserting the approximation (\ref{KK1}) in the evaluation formula for the call option, after some manipulations one can approximate  the option's price as
\begin{equation}
\label{KK2}
\begin{aligned}
u^{KK}(x,r,T;\rho) =&  \e^x \cN(d_1)- \e^{\kappa-\int_0^T \varphi(s) ds} \cN(d_2)  \\
+ & \eta C_1 \Big [d_2 \e^x \cN'(d_1)-d_1  \e^{\kappa-\int_0^T \varphi(s) ds} \cN'(d_2) \Big ]
\end{aligned}
\end{equation}
where  
$$
\begin{aligned}
C_1=& -\frac{\rho}{\sigma T} \frac{2 \sqrt{\theta} \big [(1+2 \e^{\gamma T}) \sqrt{r}-3\gamma_K\big]+\big [r-\theta (1+2 e^{\gamma T})\big] \lambda_K}{2 \e^{\gamma T} \gamma^2 \sqrt{\theta}},\\
d_1 =&\frac{x-\kappa+\theta T+(r-\theta) (1-\e^{-\gamma T})/\gamma+\sigma^2 T/ 2}{\sqrt{\sigma^2 T}}, \ \ d_2=d_1-\sigma \sqrt{T},
\end{aligned}
$$
being $\gamma_K=\e^{\gamma T/2} \sqrt{r-\theta (1-\e^{\gamma T})}$ and $\lambda_K=\log\left(\frac{ (\sqrt{r}+\sqrt{\theta})^2}{r-\theta (1-2 \e^{\gamma T})+2 \gamma_K \sqrt{\theta})} \right)$.

\subsection{Numerical results}

We compare the results of the different approximations with the benchmark Monte Carlo method, applied to the price (\ref{uBS_rate}). In particular this means that we only have to simulate the rate process to get samples from $d_1(\rho)$ and $d_2(\rho)$. The simulation was implemented by means of the Euler discretization with full truncation algorithm (see \cite{LKVD}). In our numerical experiments we generated $M=10^6$ sample paths with a time step discretization equal to $10^{-3}$ for all the maturities. All the algorithms were implemented in MatLab (R2019b) and ran on an Intel Core i7 2.40GHZ with 8GB RAM, by using the available building-in functions, in particular for the computation of all the integrals involved. The average time to compute one price was (in secs) $32.1$ (MC), $0.055$ (GO), $0.005$ (KK) and $0.009$ (MM).

We chose different set of parameters $(\kappa, \theta, \eta)$ and volatility scenarios: a low volatility $\sigma_L=0.2$ and a high volatility $\sigma_L=0.4$; hence we varied the correlation $\rho$, the rate volatility $\eta$ and the maturity of the contract $T$. The initial price of the underlying was set to $100$ as well as the strike price $K$. Numerical results are summarized in Tables (\ref{Tab1}) - (\ref{Tab4eta}). 
At least in the CIR model, the numerical results show that the MM method produces  the best approximations with respect to the benchmark Monte Carlo evaluation in most scenarios.

\newpage


\begin{table}
\centering
\begin{tabular}{c|ccccccc}
\hline
$\rho$ & -0.9 & -0.6 & -0.3 & 0.0 & 0.3 & 0.6 & 0.9 \\ \hline
& & & & Prices & & & \\ \hline
   MC&8.1543&8.1799&8.2055&8.2314&8.2574&8.2832&8.3085 \\
   & (0.0225)  &  (0.0137) &   (0.0064)  &  (0.0003) &   (0.0069)  &  (0.0142) &  (0.0230) \\ \hline
   GO&8.1192&8.1568&8.1943&8.2315&8.2686&8.3055&8.3423 \\
   KK&8.1361&8.1677&8.1993&8.2309&8.2625&8.2941&8.3258 \\
   MM& 8.146&8.1745&8.2029&8.2313&8.2595&8.2877&8.3157 \\ \hline
    & & & & Errors & & & \\ \hline
    GO& 0.0351& 0.0231& 0.0113&-0.0001& -0.0113&-0.0223& -0.0338 \\
    KK& 0.0182&0.0121&0.0062& 0.0005&-0.0052&-0.0109& -0.0172\\
    MM&0.0083&0.0053&0.0026& 0.0001&-0.0022&-0.0045&-0.0072\\

     & & & & Rel. Err. & & & \\ \hline
    GO&0.0043& 0.0028& 0.0013&1.4e-05& 0.0014& 0.0027& 0.0041\\
    KK& 0.0022& 0.0015&0.0008&5.9e-05&0.0006& 0.0013&  0.0021\\
    MM&0.0010&0.00065&0.0003&1.7e-05&0.0003&0.0005&0.0009\\

\end{tabular}
\caption{\textit{Results of the approximations for the parameters $\kappa=0.6$, $\theta=0.02$, $\eta=0.1$, $r_0=0.001$ and $\sigma_L$. The time to maturity is $T=1$ and $K=100$. In parenthesis the confidence interval of the Monte Carlo (MC) estimates. The error is defined as the difference between the MC price and the related approximation.}}
\label{Tab1}
\end{table}

\begin{table}
\centering
\begin{tabular}{c|ccccccc}
\hline
$\rho$ & -0.9 & -0.6 & -0.3 & 0.0 & 0.3 & 0.6 & 0.9 \\ \hline
& & & & Prices & & & \\ \hline
   MC&19.8443&20.1287&20.4125&20.6936&20.9705&21.2425&21.5086 \\
  &  (0.0595) &   (0.0351) &   (0.0153) &  (0.0026) & (0.0202) &  (0.0404) & (0.0649) \\ \hline
   GO&19.6375&19.9974&20.3492&20.6936&21.0308&21.3614&21.6856\\
   KK&19.7487&20.0582&20.3678&20.6773&20.9869&21.2964& 21.606\\
   MM&19.7747& 20.085&20.3892&20.6875& 20.981& 21.269&21.5522\\ \hline
    & & & & Errors & & & \\ \hline
    GO& 0.2067& 0.1313&0.0632&5.2e-07&-0.0603& -0.1189& -0.1769 \\
    KK&0.0956&0.0705&0.0447&  0.0162&-0.0164&-0.0540&-0.0973 \\
    MM&0.0695&0.0436&0.0232& 0.0061&-0.0104&-0.0265&-0.0435\\ \hline
     & & & & Rel. Err. & & & \\ \hline
    GO& 0.0104&0.0065&0.0031&2.5e-08& 0.0029&0.0056&0.0082\\
    KK&0.0048&0.0035&0.0022&0.0008&0.0008&0.0025&0.0045\\
    MM&0.0035&0.0022&0.0011&0.0003&0.0005&0.0012&0.0020\\
\end{tabular}
\caption{\textit{Results of the approximations for the parameters $\kappa=0.6$, $\theta=0.02$, $\eta=0.1$, $r_0=0.001$ and $\sigma_L$. The time to maturity is $T=5$ and $K=100$. In parenthesis the confidence interval of the Monte Carlo (MC) estimates. The error is defined as the difference between the MC price and the related approximation.}}
\label{Tab2}
\end{table}

\begin{table}
\centering
\begin{tabular}{c|ccccccc}
\hline
$\rho$ & -0.9 & -0.6 & -0.3 & 0.0 & 0.3 & 0.6 & 0.9 \\ \hline
& & & & Prices & & & \\ \hline
    MC&16.0337&16.0533& 16.073&16.0933&16.1141&16.1351& 16.156 \\
         & (0.0504) & (0.0301) & (0.0139) & (0.0002) & (0.0144) & (0.0306) & (0.0509) \\ \hline
    GO&15.9831&16.0199&16.0567&16.0934&16.1300&16.1665& 16.2030\\
    KK&15.9997&16.0309& 16.062&16.0932&16.1243&16.1555&16.1866\\
    MM&16.0094&16.0374&16.0654&16.0933&16.1211&16.1489&16.1767\\ \hline
    & & & & Errors & & & \\ \hline
    GO&0.0506&0.0333& 0.0162&-0.0001& -0.0159&-0.0314&-0.0469 \\
    KK&0.0339& 0.0224& 0.0109& 0.0001& -0.0102&-0.0203&-0.0306\\
    MM&0.0242& 0.0159&0.0076& 1.8e-05&-0.0071&-0.0138&-0.0207\\ \hline
     & & & & Rel. Err. & & & \\ \hline
    GO& 0.0032& 0.0021& 0.0010&7.0e-06&0.0010& 0.0019&0.0029\\
    KK&0.0021& 0.0013&0.0007&6.9e-06&0.0006& 0.0012&0.0019\\
    MM&0.0015&0.0009& 0.0005& 1.1e-06&0.0004&0.0009&0.0013\\ \hline
\end{tabular}
\caption{\textit{Results of the approximations for the parameters $\kappa=0.6$, $\theta=0.02$, $\eta=0.1$, $r_0=0.001$ and $\sigma_H$. The time to maturity is $T=1$ and $K=100$. In parenthesis the confidence interval of the Monte Carlo (MC) estimates. The error is defined as the difference between the MC price and the related approximation.}}
\label{Tab3}
\end{table}

\begin{table}
\centering
\begin{tabular}{c|ccccccc}
\hline
$\rho$ & -0.9 & -0.6 & -0.3 & 0.0 & 0.3 & 0.6 & 0.9 \\ \hline
& & & & Prices & & & \\ \hline
    MC&36.1379&36.3566&36.5912&36.8358&37.0875&37.3439&37.6015\\
    & (0.0405) & (0.0102) & (0.0141) & (0.0008) & (0.0163) & (0.0324) & (0.0571) \\ \hline
    GO&35.8574&36.1877&36.5138&36.8358& 37.154&37.4683&37.7789\\
    KK&35.9641&36.2539&36.5437&36.8335&37.1233&37.4132& 37.703\\
    MM&35.9876&36.2725&36.5543&36.8329&37.1089&37.3819& 37.6520\\ \hline
    & & & & Errors & & & \\ \hline
    GO&0.2805& 0.1690& 0.0774&2.2e-06&-0.0664& -0.1244&-0.1773 \\
    KK&0.1739& 0.1028&0.04748& 0.0023&-0.0358&-0.0693&-0.1014 \\
    MM&0.1504&0.0842&0.0368& 0.0029&-0.0214&-0.0379&-0.0504 \\ \hline    
    & & & & Rel. Err. & & & \\ \hline
    GO& 0.0078&0.0045&0.0021&6.1e-08& 0.0018&0.0033&0.0047\\
    KK&0.0048&0.0028&0.0013&6.2e-05&0.0010&0.0018&0.0027\\
    MM& 0.0041&0.0023& 0.0010&7.8e-05&0.0006&0.0010&0.0013\\
\end{tabular}
\caption{\textit{Results of the approximations for the parameters $\kappa=0.6$, $\theta=0.02$, $\eta=0.1$, $r_0=0.001$ and $\sigma_H$. The time to maturity is $T=5$ and $K=100$. In parenthesis the confidence interval of the Monte Carlo (MC) estimates. The error is defined as the difference between the MC price and the related approximation.}}
\label{Tab4}
\end{table}
\begin{table}
\centering
\begin{tabular}{c|ccccccc}
\hline
$\eta$ & 0.001 & 0.02 & 0.04 & 0.06 & 0.08 & 0.1 & 0.12 \\ \hline
& & & & Prices & & & \\ \hline
MC& 8.387& 8.392&8.3972&8.4025&8.4077&8.4128&8.4179 \\
 & (0.0044)  &  (0.0045) &  (0.0045) &  (0.0046) &  (0.0047) &  (0.0047) & (0.0048)\\
GO&  8.39&8.3963&8.4029&8.4095& 8.416&8.4224&8.4286 \\
KK&8.3899&8.3949&8.4001&8.4053&8.4105&8.4158& 8.421\\
MM&8.3899&8.3944&8.3992&8.4039&8.4086&8.4132&8.4177\\ \hline
    & & & & Errors & & & \\ \hline
    GO&-0.0030&-0.0043&-0.0056&-0.0070& -0.0083& -0.0096& -0.0107 \\
    KK&-0.0029&-0.0029&-0.0029&-0.0029& -0.0029& -0.0027&-0.0031 \\
    MM&-0.0029&-0.0024&-0.0019&-0.0014&-0.0009&-0.0004&0.0002\\ \hline
    & & & & Rel. Err. & & & \\ \hline
    GO&0.0004&0.0005&0.0007&0.0008&0.0010& 0.0011&  0.0012 \\
    KK&0.0003& 0.0003&0.0003&0.0003&0.0003&0.0004&0.0004 \\
    MM&0.0003&0.0003&0.0002&0.0002&0.0001& 4.8e-05&2.1e-05\\
\end{tabular}
\caption{\textit{Results of the approximations for the parameters $\kappa=0.58$, $\theta=0.0345$, $\rho=0.2$, $r_0=0.001$ and $\sigma_L$. The time to maturity is $T=1$ and $K=100$. In parenthesis the confidence interval of the Monte Carlo (MC) estimates. The error is defined as the difference between the MC price and the related approximation.}}
\label{Tab1eta}
\end{table}

\begin{table}
\centering
\begin{tabular}{c|ccccccc}
\hline
$\eta$ & 0.001 & 0.02 & 0.04 & 0.06 & 0.08 & 0.1 & 0.12 \\ \hline
& & & & Prices & & & \\ \hline
    MC&16.2303&16.2352&16.2403&16.2454&16.2504&16.2553&16.2601\\
     & (0.0093) &  (0.0094) & (0.0094)  &  (0.0095) &  (0.0096) &  (0.0096) &  (0.0097) \\
    GO&16.2366&16.2427&16.2492&16.2556&16.2618& 16.268&16.2738\\
    KK&16.2365&16.2414&16.2466&16.2517&16.2569& 16.262&16.2671\\
    MM&16.2365&16.2409&16.2456&16.2502&16.2547&16.2591&16.2634 \\ \hline
    & & & & Errors & & & \\ \hline
    GO&-0.0063&-0.0075&-0.0089& -0.0102& -0.0115& -0.0127& -0.0138\\
    KK&-0.0062& -0.0062&-0.0063&-0.0063&-0.0065&-0.0067&-0.0071\\
    MM&-0.0062&-0.0057&-0.0053&-0.0048& -0.0043&-0.0038&-0.0033  \\ \hline
    & & & & Rel. Err. & & & \\ \hline
    GO&0.0004&0.0005&0.0005&0.0006&0.0007&0.0008&0.0008\\
    KK& 0.0004&0.0004&0.0004&0.0004&0.0004& 0.0004&0.0004\\
    MM& 0.0004&0.0004&0.0003&0.0003&0.0003&0.0002&0.0002\\
\end{tabular}
\caption{\textit{Results of the approximations for the parameters $\kappa=0.58$, $\theta=0.0345$, $\rho=0.2$, $r_0=0.001$ and $\sigma_H$. The time to maturity is $T=1$ and $K=100$. In parenthesis the confidence interval of the Monte Carlo (MC) estimates. The error is defined as the difference between the MC price and the related approximation.}}
\label{Tab2eta}
\end{table}

\begin{table}
\centering
\begin{tabular}{c|ccccccc}
\hline
$\eta$ & 0.001 & 0.02 & 0.04 & 0.06 & 0.08 & 0.1 & 0.12 \\ \hline
& & & & Prices & & & \\ \hline
    MC&22.8358&22.8854&22.9389&22.9928&23.0467&23.0999&23.1516\\
    & (0.0122) &  (0.0128)  &  (0.0135)  &  (0.0142) &   (0.0150) &   (0.0157) & (0.0164) \\
    GO&22.8415&22.9009&22.9645&23.0286&23.0925&23.1554&23.2165\\
    KK& 22.841&22.8902& 22.942&22.9939&23.0457&23.0975&23.1493\\
    MM& 22.841&22.8918&22.9461&23.0007&23.0546&23.1069&23.1565 \\ \hline
    & & & & Errors & & & \\ \hline
    GO&-0.0057& -0.0154& -0.0257&  -0.0358& -0.0458& -0.0556& -0.0649\\
    KK&-0.0052&-0.0048&-0.0032&-0.0010&  0.0010& 0.0023& 0.0023\\
    MM&-0.0052&-0.0064&-0.0073&-0.0078&-0.0079&-0.0070&-0.0050\\ \hline
    & & & & Rel. Err. & & & \\ \hline
    GO&0.0002&0.0007& 0.0011& 0.0016& 0.0020& 0.0024& 0.0028\\
    KK&0.0002& 0.0002&0.0001&4.4e-05&4.4e-05&0.0001&0.0001\\
    MM&0.0002&0.0002&0.0003&0.0003&0.0003&0.0003&0.0002\\
\end{tabular}
\caption{\textit{Results of the approximations for the parameters $\kappa=0.58$, $\theta=0.0345$, $\rho=0.2$, $r_0=0.001$ and $\sigma_L$. The time to maturity is $T=5$ and $K=100$. In parenthesis the confidence interval of the Monte Carlo (MC) estimates. The error is defined as the difference between the MC price and the related approximation.}}
\label{Tab3eta}
\end{table}

\begin{table}
\centering
\begin{tabular}{c|ccccccc}
\hline
$\eta$ & 0.001 & 0.02 & 0.04 & 0.06 & 0.08 & 0.1 & 0.12 \\ \hline
& & & & Prices & & & \\ \hline
    MC&38.4209&38.4676&38.5166&38.5649&38.6121&38.6578&38.7014\\
    & (0.0257) &   (0.0262) &   (0.0267) &   (0.0272)  &  (0.0278) &   (0.0283)  &  (0.0289) \\
    GO&38.4364&38.4922&38.5505& 38.608&38.6639&38.7179&38.7692\\
    KK&38.4359&38.4828&38.5321&38.5815&38.6309&38.6802&38.7296\\
    MM& 38.436&38.4836&38.5333& 38.582& 38.629&38.6737&38.7154\\ \hline
    & & & & Errors & & & \\ \hline
    GO&-0.0155&-0.0246&-0.0340&-0.0431&-0.0518&-0.0601&-0.0678\\
    KK&-0.0150&-0.0152&-0.0156&-0.0166&-0.0187&-0.0224&-0.0281\\
    MM&-0.0150&-0.0160&-0.0167&-0.0171&-0.0169&-0.0159&-0.0140\\ \hline
    & & & & Rel. Err. & & & \\ \hline
    GO&0.0004&0.0006&0.0009&0.0011&0.0013& 0.0016&  0.0018\\
    KK&0.0004&0.0004&0.0004&0.0004&0.0005&0.0006&0.0007\\
    MM&0.0004&0.0004&0.0004&0.0004&0.0004&0.0004&0.0004\\
\end{tabular}
\caption{\textit{Results of the approximations for the parameters $\kappa=0.58$, $\theta=0.0345$, $\rho=0.2$, $r_0=0.001$ and $\sigma_H$. The time to maturity is $T=5$ and $K=100$. In parenthesis the confidence interval of the Monte Carlo (MC) estimates. The error is defined as the difference between the MC price and the related approximation.}}
\label{Tab4eta}
\end{table}


\begin{thebibliography}{00}



\bibitem {AJ92} Amin, K. and R. Jarrow,  \textit{Pricing options on risky assets in a stochastic interest rate economy}, Mathematical Finance 2, (1992) 217 -- 237.


\bibitem{BCC97} Bakshi, G., Cao, C., Chen, Z., \textit{Empirical performance of alternative option pricing models}, The Journal of finance, 52(5), (1997) 2003--2049.

\bibitem{BCC00} Bakshi, G., Cao, C., Chen, Z.  \textit{Pricing and hedging long-term options}, Journal of econometrics, 94(1-2), (2000) 277--318.

\bibitem{B} T. Bjork, \textit{Arbitrage Theory in Continuous Time}, Oxford University Press 2009.

\bibitem{BM13} D. Brigo, F. Mercurio, \textit{Interest Rates Models}, Springer 2013.

\bibitem{BR18} D. Brigo, F. Vrins, \textit{Disentangling wrong-way risk: pricing credit valuation adjustment via change of measures}, European Journal of Operational Research 269 (2018) 1154--1164.

\bibitem{DPS00} D. Duffie, J. Pan, K. J. Singleton, \textit{Tranform analysis and asset pricing for affine jum-diffusions}, Econometrica, Vol. 68, No. 6, November, (2000), 1343--1376.


\bibitem{Dufr} D. Dufresne, \textit{The integrated square root process}, Research Paper, University of Melbourne, (2001).

\bibitem{GO} L.A. Grzelak, C.W. Oosterlee, \textit{On the Heston model with stochastic interest rates}, SIAM Journal on Financial Mathematics 2 (1) (2011) 255-286.

\bibitem{KS} I. Karatzas, S: Shreve \textit{Brownian Motion and Stochastic Calculus}, Graduate Texts in Mathematics 113, Springer-Verlag New Yprk (1998)..

\bibitem{KK99} Y. J. Kim, N. Kunimoto, \textit{Pricing Options under Stochastic Interest Rates: A New Approach.} Asia-Pacific Financial Markets, 6, 49--70 (1999).

\bibitem{Kim02} Y.J. Kim, \textit{Option Pricing under Stochastic Interest Rates: An Empirical Investigation}, Asia-Pacific Financial Markets 9, (2002) 23 -- 44.

\bibitem{Lee04} R. W. Lee, \textit{Option Pricing by Transform Methods: Extensions, Unification, and Error Control}, Journal of Computational Finance, 7, 51--86, (2004). 

\bibitem{LKVD} R. Lord, R. Koekkoek, D. Van DijK, \textit{A comparison of biased simulation schemes for the stochastic volatility models}, Quantitative Finance, 10 (2) (2010) 177--194.

\bibitem{M73} Merton, R., \textit{The theory of rational option pricing}, Bell J. Econom. Managt Sci. 4, (1973) 14 -- 183.

\bibitem{Pr} P. Protter, \textit{Stochastic Integration and Differential Equations}, Stochastic Modelling and Applied Probability, 21, Springer-Verlag Berlin Heidelberg (2005).

\bibitem{Rab89} Rabinovitch, R. \textit{Pricing stock and bond options when the default-free rate is stochatic}, J. Financ. Quantitat. Anal. 24, (1989) 447 -- 457.

\bibitem{RS09} A. Ramponi, S. Scarlatti, S. \textit{Option pricing in a hidden Markov model of the short rate with application to risky debt evaluation}, Int. J. Risk Assessment and
Management, Vol. 11, (1/2), (2009)  88--103.

\bibitem{Rin95} Rindell, K. \textit{Pricing of index options when interest rates are stochastic: an empirical test}, J. Banking Finance 19, (1995) 785 -- 802.

\bibitem{STV93} Shimko D.C., Tejima N., Van Deventer D.  \textit{The pricing of risky debts when interest rates are stochastic}, Journal of Fixed Income Vol. 3, No.2, (1993), pp. 58--65.

\bibitem{ZA81} S. Zacks, \textit{Parametric Statistical Inference}, Pergamon Press 1981.

\end{thebibliography}
\end{document}